# Planar topological Hall effect in a hexagonal ferromagnetic $Fe_5Sn_3$ single crystal

Hang Li[1], Bei Ding[1], Jie Chen[1,3], Zefang Li[1,2], Xuekui Xi[1], Guangheng Wu[1], and Wenhong Wang[1,3]*

[1]Beijing National Laboratory for Condensed Matter Physics, Institute of Physics, Chinese Academy of Sciences, Beijing 100190, China
[2]University of Chinese Academy of Sciences, Beijing 100049, China
[3] Songshan Lake Materials Laboratory, Dongguan, Guangdong 523808, China

*Corresponding author. Email: wenhong.wang@iphy.ac.cn

## Abstract

The planar topological Hall effect (PTHE), appeared when the magnetic field tended to be along the current, is believed to result from the real-space Berry curvature of the spin spiral structure and has been experimentally observed in skyrmion-hosting materials. In this paper, we report an experimental observation of the PTHE in a hexagonal ferromagnetic $Fe_5Sn_3$ single crystal. With a current along the $c$ axis of $Fe_5Sn_3$, the transverse resistivity curves exhibited obvious peaks near the saturation field as the magnetic field rotated to the current and appeared more obvious with increasing temperature, which was related to the noncoplanar spin structure in $Fe_5Sn_3$. This spin structure induced nonzero scalar spin chirality, which acted as fictitious magnetic fields to conduction electrons and contributed the additional transverse signal. These findings deepen the understanding of the interaction between conduction electrons and complex magnetic structures and are instructive for the design of next-generation spintronic devices.

Topological magnetic structure, like skyrmion, have greatly inspired the development of spintronics[1]. Moreover, the fascinating electromagnetic phenomena from the interaction of conduction electrons and topological magnetic structures also promoted the practicality of next-generation spintronic devices [2]. Among these electromagnetic phenomena, the Hall effect is one of the easiest to measure and be applied in devices [3, 4]. In noncoplanar magnetic materials, in addition to the normal [5] and anomalous Hall effects [6], an additional Hall contribution, named the topological Hall effect (THE), also becomes the hotspot of the research, which is induced by the spatial distribution of spin [7, 8]. In these magnetic materials, the three nearest spins generate nonzero scalar spin chirality, which acts as fictitious magnetic fields to conduction electrons and induces the so called THE [7]. Thus far, the THE has been observed in many materials, such as artificial magnetic nanostructures [9-12], frustrated magnetic materials [7], noncollinear antiferromagnetic materials [13-15] and skyrmion-hosting materials [4, 16-19]. Furthermore, the THE even becomes, to some degree, a fingerprint of the topological magnetic structure.

In addition to the THE, an additional Hall signal, called the planar topological Hall effect (PTHE), wherein the Hall voltage and electric and magnetic fields are coplanar, has very recently been observed in skyrmion-hosting materials, like $Fe_3GeTe_2$ [20]. This signal is directly related to the in-plane component of topological magnetic textures with an external magnetic field applied in the plane of the charge current [20, 21]. Additionally, theoretical work suggests that the PTHE can also originate from the real-space Berry curvature of the conical spin spiral texture and can appear in a variety of topological materials [22]. Therefore, the PTHE also becomes an important characteristic to search the topological magnetic materials.

In this paper, we report an experimental observation of the PTHE with current $I$ along the $c$ axis in a hexagonal ferromagnetic $Fe_5Sn_3$ single crystal. $Fe_5Sn_3$ is classified as a hexagonal crystal system with space group $P6_3/mmc$ ($a = b = 4.224$ Å, $c = 5.222$ Å). As shown in Fig. 1(a), the Fe atoms are located in Fe and Fe-Sn layers, which are named Fe-I (0, 0, 0) and Fe-II (1/3, 2/3, 1/4), respectively. However, vacancies and disorder always form at the Fe-II location, impacting the physical properties of $Fe_5Sn_3$ [23, 24]. The anisotropic magnetization curves indicate that the easy and hard magnetization axes are the *b* (*a*) axis and *c* axis of $Fe_5Sn_3$, respectively [25]. For the experimental part, we measured the angle-dependent longitudinal and transverse resistivity along different axes of $Fe_5Sn_3$. First, at 300 K, when current *I* was along the *c* axis, a peak gradually appeared near the saturation field in the transverse resistivity curve as the magnetic field *H* rotated to the direction of current *I*. Moreover, the peak still existed even when *H* was parallel to *I*. Second, when *H* was parallel to *I*, peaks appeared at all measurement temperatures (5-350 K), and the amplitude of the peaks increased with temperature. Third, peaks also appeared in the longitudinal resistivity curves as *H* rotated in the *bc* plane with *I* along the *c* axis. Combined with the reported THE of $Fe_5Sn_3$ [26], we thought that the observed peaks in the transverse resistivity curves were also related to the nonzero scalar spin chirality induced by the noncoplanar magnetic structure in $Fe_5Sn_3$ [22]. These experimental phenomena can deepen the understanding of the interaction between conduction electrons and complex magnetic structures and provide design ideas for next-generation spintronic devices.

$Fe_5Sn_3$ single crystals were synthesized by the Sn self-flux method, which has been described in a previous report [27]. The isolated single crystal appearance is a needle-shaped hexagonal prism. X-ray diffraction data have confirmed that the preferred growth orientation of a single crystal is along the *c* axis and that the rectangular side face is the *bc* plane of $Fe_5Sn_3$ [27]. Since the Fe-II vacancies (as shown in Fig. 1(a)) are related to the synthesis temperature [24], to minimize the impact of Fe-II vacancies in different samples, the single crystal samples measured in this paper were from the same batch. The magnetic and electrical properties were both measured in a physical property measurement system (Quantum Design, PPMS-9T). A standard four-wire method was applied in electric transport measurements. To deduct superimposed signals introduced by wire misalignment, the longitudinal ($\rho_{lon}(H)$) and transverse ($\rho_{tra}(H)$) resistivities were both measured at positive and negative magnetic fields ($\rho_{lon}(H) = [\rho_{lon}(+H) + \rho_{lon}(-H)]/2$, $\rho_{tra}(H) = [\rho_{tra}(+H) - \rho_{tra}(-H)]/2$). All measurements were repeated on multiple samples. The band structure and partial density of states (PDOS) of $Fe_5Sn_3$ were calculated by the WIEN2K code package, which was based on density functional theory (DFT) [28].

The longitudinal resistivity curves at zero magnetic field are shown in Fig. 1(b), and all curves exhibited typical metallicity and weak anisotropy. The small residual resistance rates (RRR=$R_{300K}/R_{2K}$) implied that some defects exist in the single crystals. Figure 1(c) shows the band structure and PDOS of $Fe_5Sn_3$, which indicated that $Fe_5Sn_3$ is a typical magnetic metal and that the magnetic moments and conduction electrons are mostly contributed by *d* electrons of Fe atoms. Calculations showed that Fe-I and Fe-II atoms have different atomic magnetic moments (1.77 $\mu_B$/Fe-I and 2.47 $\mu_B$/Fe-II) and that the average magnetic moment of Fe atoms is 2.12 $\mu_B$/Fe, which is almost consistent with the experimental values [27].

A large intrinsic anomalous Hall effect [27] and an anisotropic THE [26] have been reported in $Fe_5Sn_3$. These phenomena implied that nontrivial band and magnetic structures may exist in $Fe_5Sn_3$. To further explore the magnetic structure in $Fe_5Sn_3$, as shown in Fig. 2, we plotted the transverse resistivities and magnetoresistance (MR = $[\rho_{lon}(H) - \rho_{lon}(0)]/\rho_{lon}(0)$) curves along different axes for various angles between *H* and *I*. Figures 2(a), (d) and (g) show schematic diagrams of the measurement. To clearly describe the experimental phenomena, we slightly shifted the MR curves up and down and defined the *b* (*a*) and c axes as the *x* and *z* directions, respectively. The angle between *I* and *H* was defined as $\theta$.

As shown in Figs. 2(b) and (c), at 300 K, when *I* was along the *c* axis and *H* rotated around the *b* axis, the transverse resistivity $\rho_{xz}(H)$ and MR curves showed some evident peaks when $\theta$ was less than 75°. Moreover, the shapes of the $\rho_{xz}(H)$ curves were very similar to those of the Hall resistivity curves in some skyrmionic materials with the THE [16, 29]. As shown in Fig. S1 , the critical fields of these peaks in the $\rho_{xz}(H)$ and MR curves were consistent and close to the saturation field in this direction [26], which implied that they were related to the magnetic structure of $Fe_5Sn_3$. Furthermore, the critical field of these peaks, as shown by the red points in Figs. 2(b) and (c), increased with decreasing $\theta$ because the *c* axis was the hard axis of $Fe_5Sn_3$ and the saturation field increased as *H* rotated from the direction of the easy magnetization plane (*ab* plane) to the hard axis (*c* axis) direction [25, 26]. Another interesting result was that the transverse resistivity $\rho_{xz}(H)$ was nonzero even at $\theta=0°$, which is consistent with the situation in the skyrmionic material $Fe_3GeTe_2$ [20, 30, 31]. To exclude the possibility that the nonzero signal of $\rho_{xz}(H)$ at $\theta=0°$ arose from angular deviation, we also measured the $\rho_{xz}(H)$ curves over a small angle range ($-5° \leq \theta \geq 5°$), as shown in Fig. S2. The unequal absolute values of $\rho_{xz}(H)$ at $\theta=-5°$ and $5°$ did indicate the existence of angular deviation, but the $\rho_{xz}(H)$ values remained nonzero at all measurement angles, and a sharp peak appeared in the $\rho_{xz}(H)$ curve in the low field region at $\theta=-2°$, which is shown in the inset of Fig. S2. As shown in Fig. 2(c), the MR curves also exhibited a more complex relationship with the magnetic field below the saturation field at $\theta=0°$, which meant that these peaks in the $\rho_{xz}(H)$ and MR curves were related to the magnetic structure in $Fe_5Sn_3$.

Figures 2(d-f) show the situation for *I* along the *b* (*a*) axis and *H* rotated around the *c* axis. The values of transverse resistivity $\rho_{zx}(H)$ gradually decreased as the angle $\theta$ decreased and almost disappeared at $\theta=0°$. The $\rho_{zx}(H)$ and MR curves did not show any obvious peaks at any measured angle, similar to the situation of $Fe_3Sn_2$ with *I* along the *ab* plane and *H* rotated from the *c* axis to the *ab* plane (as shown in Fig. S3(a)). The difference between $Fe_5Sn_3$ and $Fe_3Sn_2$ was that the MR of $Fe_5Sn_3$ did not exhibit obvious variation, as shown in Figs. 2(f) and S3(b). A previous article reported that skyrmions were only generated when *H* tended to be along the *c* axis in $Fe_3Sn_2$ [32], so almost no additional Hall effect was introduced by skyrmions in $Fe_3Sn_2$ with *H* along the *ab* plane ($\theta=0°$). We thought that the change in $\rho_{zx}(H)$ and MR in $Fe_5Sn_3$ with $\theta$ meant that the magnetic structure in $Fe_5Sn_3$ tended to form a coplanar structure as *H* rotated around the *c* axis. The *b* (*a*) axis is the easy axis of $Fe_5Sn_3$, so the spin is easier to turn into the direction of the external magnetic field when *H* is parallel to the *ab* plane. In this case, no local spin chirality was induced by the noncoplanar magnetic structure when *H* rotated around the *c* axis. As shown in Fig. 2(h), the transverse resistivity $\rho_{yx}(H)$ also appeared to have a nonzero value and small peaks near the saturation field at $\theta=0°$ as *H* rotated from the *c* to *a* axis. The MR,

as shown in Fig. 2(i), was similar to the situation in Fig. 2(f) because the current was along equivalent axes in these two samples. The angle-dependent transverse resistivity at different fields and temperatures is shown in Fig. S4. When $I$ was along the $c$ axis, as shown in Figs. S4(a-c), the transverse resistivity $\rho_{xz}(H)$ curves presented as a sine function of $\theta$, and the amplitude increased with $H$. When $I$ was along the $b$ axis, as shown in Figs. S4(d-f), the transverse resistivity $\rho_{zx}(H)$ curves were closer to a sine function as $H$ increased. At a low field, in the process of $H$ turning toward $I$, the curves became nearly linearly related to $\theta$. When $I$ was along the $ab$ plane, as shown in Figs. S4(g-i), the transverse resistivity $\rho_{yx}(H)$ curves showed an abrupt turn near the position where $H$ was perpendicular to $I$, similar to the situation of $Fe_3GeTe_2$ [20]. Moreover, the abrupt turn was gradually suppressed as $H$ increased. The abrupt change in the $\rho_{yx}(H)$ curves was related to the large magnetic anisotropy of $Fe_5Sn_3$.

As shown in Fig. 3, we also measured the transverse resistivity and MR at $\theta=0°$ from 5 to 350 K for $H$ and $I$ along the $c$ axis. As shown in Figs. 3(a) and (b), peaks in the transverse resistivity $\rho_{xz}(H)$ and MR curves appeared at all measured temperatures, and the critical field (as shown by the black dashed line in Fig. 3(a)) gradually increased as the temperature decreased because the easy axis of $Fe_5Sn_3$ changed from the $c$ axis to the $ab$ plane as the temperature decreased [27]. Furthermore, as shown in Figs. 3(c) and (d), we also extracted the planar topological Hall resistivity $\rho_{xz}^{PT}(H)$ from $\rho_{xz}(H)$ by the method described for $Fe_3GeTe_2$ [20] and plotted the contour map of $\rho_{xz}^{PT}(H)$. Figure 3(d) clearly shows that the PTHE is very obvious near the saturation field.

When $H$ and $I$ were along the $b$ axis, as shown in Figs. S5(a-c), no obvious turning points were observed in the transverse resistivity and MR curves. However, when $H$ and $I$ were along the $a$ axis, as shown in Figs. S5(d-f), the transverse resistivity $\rho_{yx}(H)$ was larger than that for the other two directions and exhibited small peaks near the saturation field at all measured temperatures. Moreover, the MR curves were still similar to the situation in Fig. S5(c). In all three cases above, the MR curves all changed their sign as the temperature increased from 5 to 350 K, which was associated with the transformation of the magnetic structure in $Fe_5Sn_3$ with temperature [26, 27]. By comparing these measurement curves, the curves measured with *H* and *I* along the *c* axis showed more obvious changes than those for the other two directions, which implies that the magnetic structure is more complex for *H* along the *c* axis.

Figure 4 shows the transverse resistivity $\rho_{xz}(H)$ and MR curves with *H* rotated in the *bc* plane at 300 K. As shown in Fig. 4(a), the transverse resistivity $\rho_{xz}(H)$ curves also exhibited peaks near the saturation field, and the magnitude of the peaks increased as $\theta$ decreased. The MR curves, as shown in Fig. 4(b), also exhibited some changes near the corresponding critical field. These critical fields of the $\rho_{xz}(H)$ and MR curves also increased as $\theta$ decreased. We further measured the angle-dependent $\rho_{xz}(H)$ and longitudinal resistivity $\rho_{zz}(\theta)$ curves, which are shown in Figs. 4(c) and (d). At 0.5 and 1 T, the angle-dependent $\rho_{xz}(H)$ curves, as shown in Fig. 4(c), also exhibited an abrupt change when *H* was near the direction of *I*, very similar to the situation in Fig. S4(g), because the plane of the magnetic field rotation in these two samples is equivalent. To clearly show the change in longitudinal resistivity with $\theta$, we plotted the difference curves between $\rho_{zz}(\theta)$ and $\rho_{zz}(0)$, as shown in Fig. 4(d). At 300 K, the curves gradually exhibited twofold symmetry as *H* increased. More detailed measurements of the angle-dependent $\rho_{xz}(H)$ and longitudinal resistivity $\rho_{zz}(\theta)$ curves are shown in Fig. S6. At 0.5 T, an abrupt change in the $\rho_{xz}(H)$ curves appeared at all measured temperatures, but it was suppressed at 5 T. Another interesting phenomenon was that the phase of the difference curves between $\rho_{zz}(\theta)$ and $\rho_{zz}(0)$ changed by 180° at 200 K, which is close to the spin reorientation temperature in $Fe_5Sn_3$ [27].

Combining all the above phenomena, the peaks present in the transverse resistivity and MR curves were more likely to appear when *H* was rotated to the *c* axis in $Fe_5Sn_3$. That may be due to the noncoplanar magnetic structure tends to appear when *H* is along the *c* axis. Similar to the situation in $Fe_{1.3}Sb$ [33], the lack of spatial inversion symmetry between Fe-I and Fe-II atoms of the nearest layers in $Fe_5Sn_3$ introduces a finite Dzyaloshinsky–Moriya interaction (DMI) whose vector direction is along the *c* axis. The competition between DMI and Heisenberg exchange interaction tends to form noncollinear magnetic structure. Considering that the *c* axis is the hard axis in $Fe_5Sn_3$, at external magnetic field along the *c* axis, the magnetic structure tends to form a conical spin spiral arrangement. In a recent theoretical work [22], except for the in-plane skyrmion tubes, the conical spin spiral also can contribute to the so-called PTHE. Therefore, we thought that the PTHE in $Fe_5Sn_3$ was related to the noncoplanar spin arrangement induced by the competition between the DMI, Heisenberg exchange interaction, and magnetocrystalline anisotropy. At proper external magnetic field, the spins arrange noncoplanar and contribute a finite local spin chirality, which induces a Berry phase and acts like a fictitious field $H_f$ to the conduction electrons, as shown in inset of Fig. 3(d), the component of fictitious field perpendicular to *I* will induce the additional transverse velocity to the conduction electrons.

In summary, we measured the angle-dependent transverse and longitudinal resistivities with *H* and *I* along different axes of $Fe_5Sn_3$. Some obvious peaks appeared in the transverse and longitudinal resistivity curves whose critical fields were consistent with each other. These additional signals showed anisotropy for *H* along different axes of $Fe_5Sn_3$ and even became more obvious when *H* and *I* tended to be coplanar. We thought that the additional signal, called the PTHE, came from the noncoplanar magnetic structure in $Fe_5Sn_3$, which was induced by the competition between the DMI, Heisenberg exchange interaction, and magnetocrystalline anisotropy. These interesting phenomena indicate that $Fe_5Sn_3$ could be a topological magnetic material candidate and also deepen the understanding of the interaction between conduction electrons and complex magnetic structures.

See supplementary materials for more details of angle dependent transverse and longitudinal resistivity curves at I along different axes of $Fe_5Sn_3$.

This work is supported by the National Key R&D Program of China (Grant Nos. 2017YFA0206303) and the National Natural Science Foundation of China (Nos.

## DATA AVAILABILITY

The data that support the findings of this study are available from the corresponding author upon reasonable request.

Figure captions

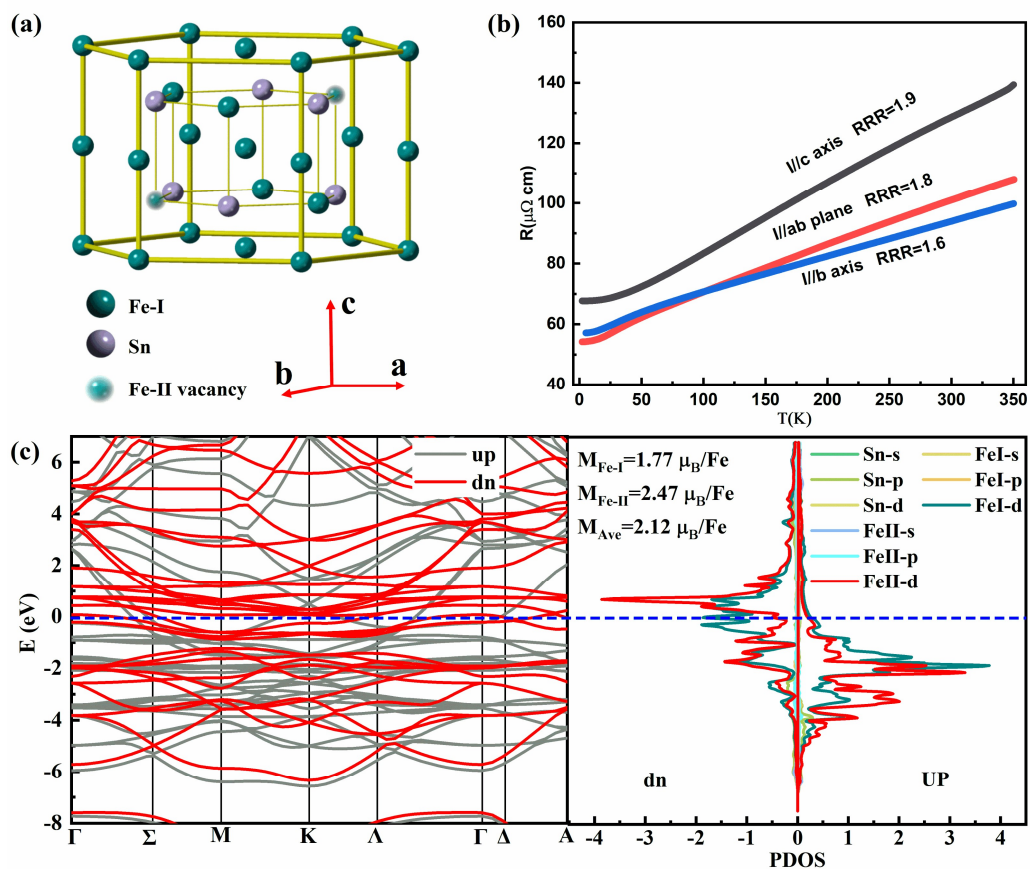

**FIG. 1** (color online). (a) Schematic diagram of the $Fe_5Sn_3$ crystal structure. (b) Resistivity curves for *I* along different directions at zero magnetic field. (c) Band structure (left) and PDOS (right) of $Fe_5Sn_3$ calculated by WIEN2K code package. The blue dashed line represents the position of the Fermi level.

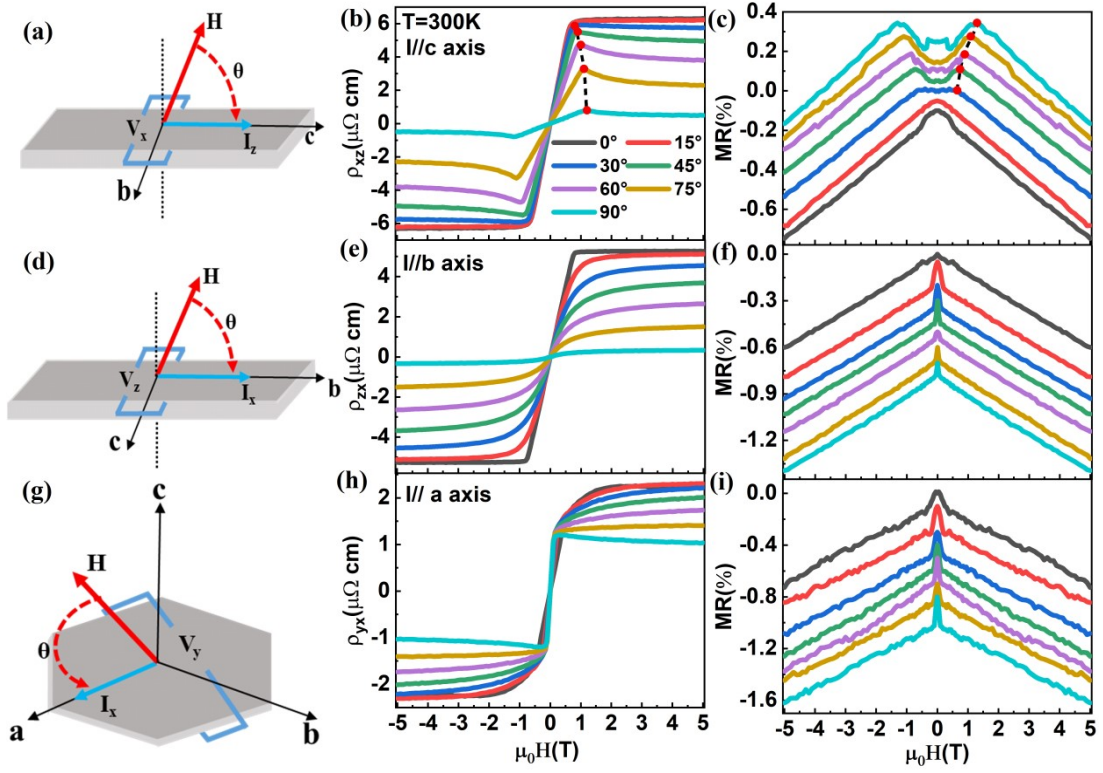

**FIG. 2** (color online). Transverse resistivity and magnetoresistance with magnetic field $H$ rotated outside the current $I$ plane at 300 K. (a-c) Situation with $H$ rotated around the $b$ axis and $I$ along the $c$ axis. The red points are the turning points of the transverse resistivity and magnetoresistance curves, which are connected with black dashed lines. (d-f) Situation with $H$ rotated around the $c$ axis and $I$ along the $b$ axis. (h-j) Situation with $H$ rotated in the $ac$ plane and $I$ along the $a$ axis. To clearly show the details of the MR curve, these MR curves are slightly shifted.

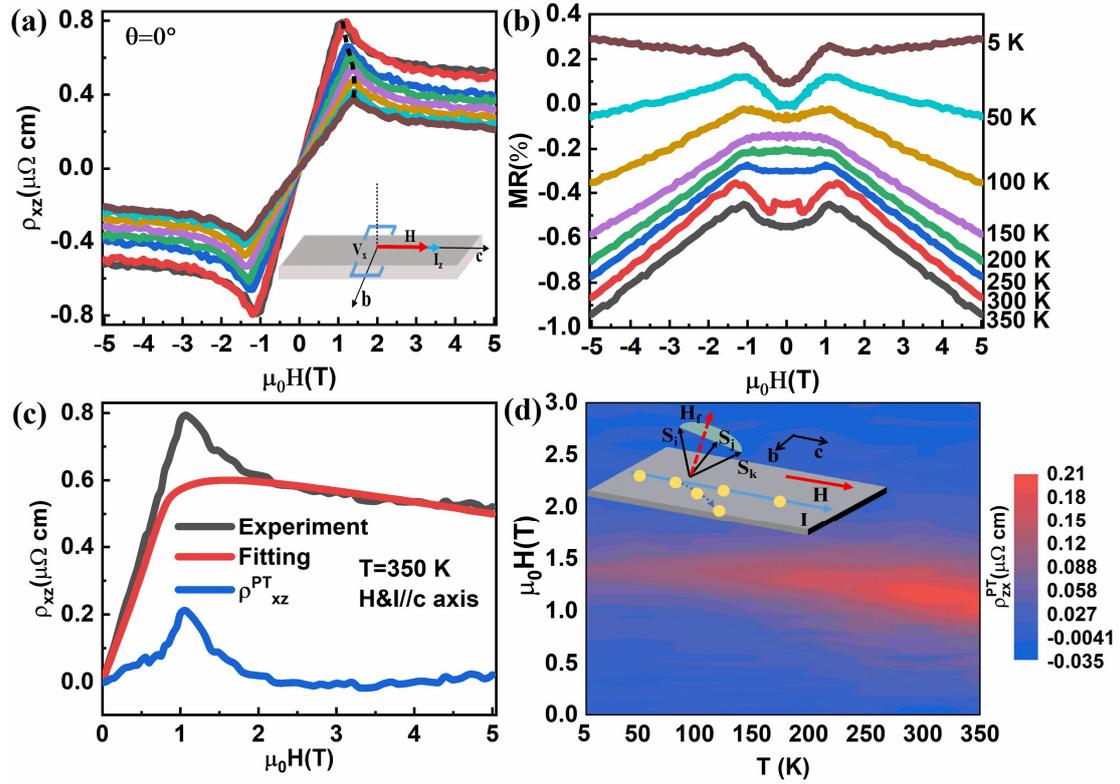

**FIG. 3** (color online). (a) Transverse resistivity and (b) magnetoresistance curves with *H* and *I* parallel to the *c* axis at various temperatures. To clearly show the details of the MR curve, these MR curves are slightly shifted. The black dashed line in (a) represents the position of peaks in the transverse resistivity. (c) Process of extracting the PTHE. The black, red and blue curves represent the experimental, fitting and PTHE curves, respectively. (d) Contour map of $\rho_{xz}^{PT}(H)$. The inset is the schematic diagram of interaction between conduction electrons and magnetic structure.

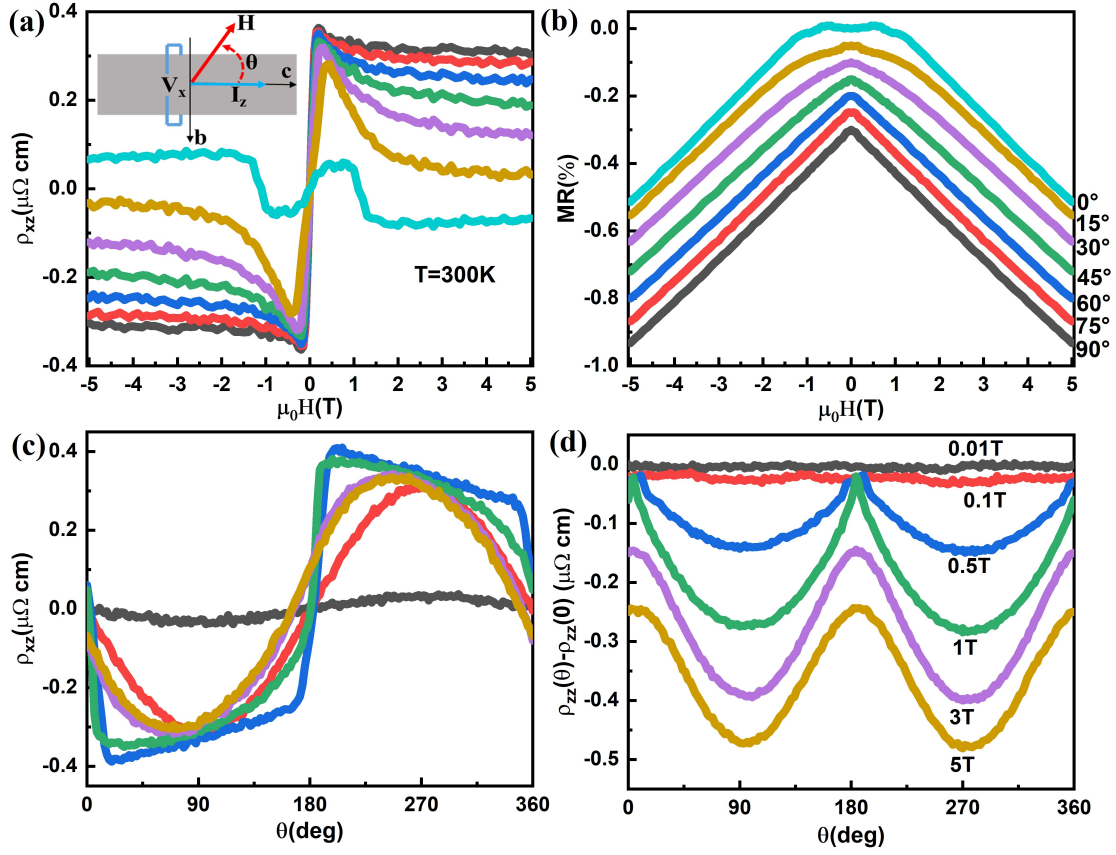

**FIG. 4** (color online). (a) Transverse resistivity and (b) magnetoresistance at different angles $\theta$ between magnetic field $H$ and current $I$ in the $bc$ plane at 300 K. The inset of (a) is a schematic diagram of the measurement process. (c) and (d) Angle-dependent $\rho_{xz}(H)$ and longitudinal resistivity $\rho_{zz}(\theta)$ difference curves at various magnetic fields.

Supplemental Materials for

# Planar topological Hall effect in a hexagonal ferromagnetic $Fe_5Sn_3$ single crystal

Hang Li[1], Bei Ding[1], Jie Chen[1,3], Zefang Li[1,2], Xuekui Xi[1], Guangheng Wu[1], and Wenhong Wang[1,3]*

[1]Beijing National Laboratory for Condensed Matter Physics, Institute of Physics, Chinese Academy of Sciences, Beijing 100190, China

[2]University of Chinese Academy of Sciences, Beijing 100049, China

[3] Songshan Lake Materials Laboratory, Dongguan, Guangdong 523808, China

*Corresponding author. Email: wenhong.wang@iphy.ac.cn

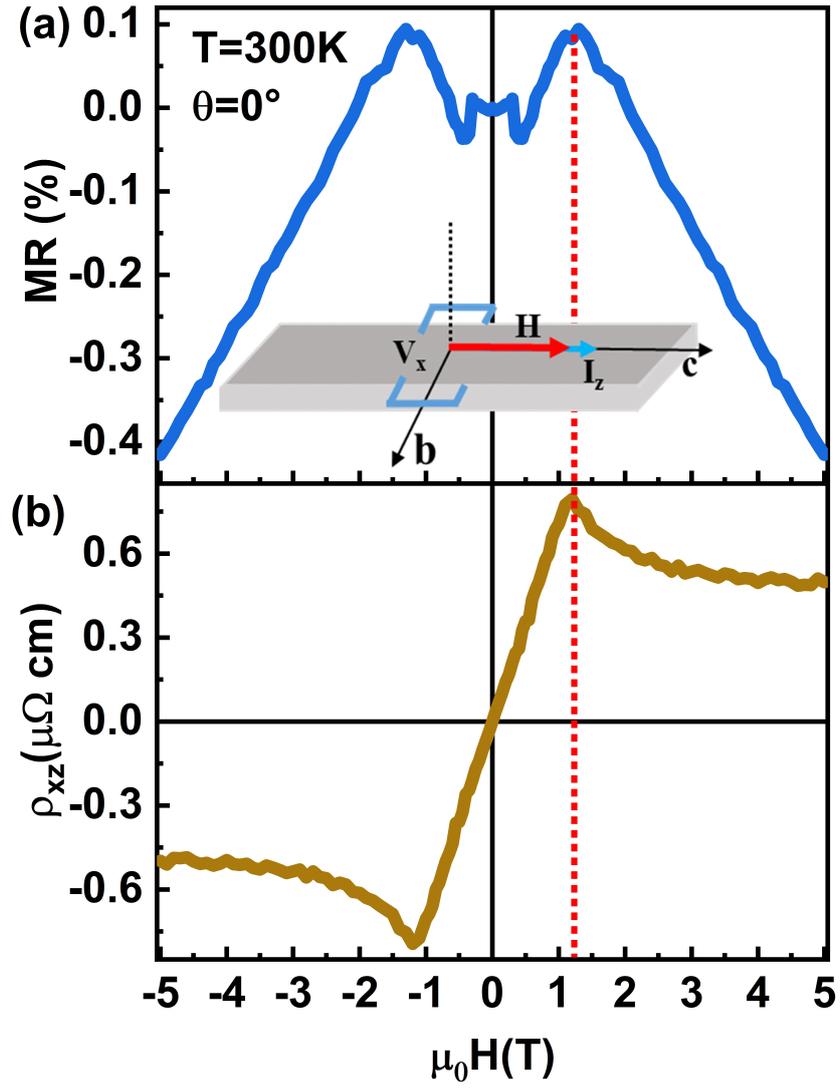

**Fig. S1.** (a) Magnetoresistance and (b) transverse resistivity $\rho_{xz}$ with $H$ and $I$ along $c$ axis at 300 K. The inset of (a) is shown the schematic diagram of measurement.

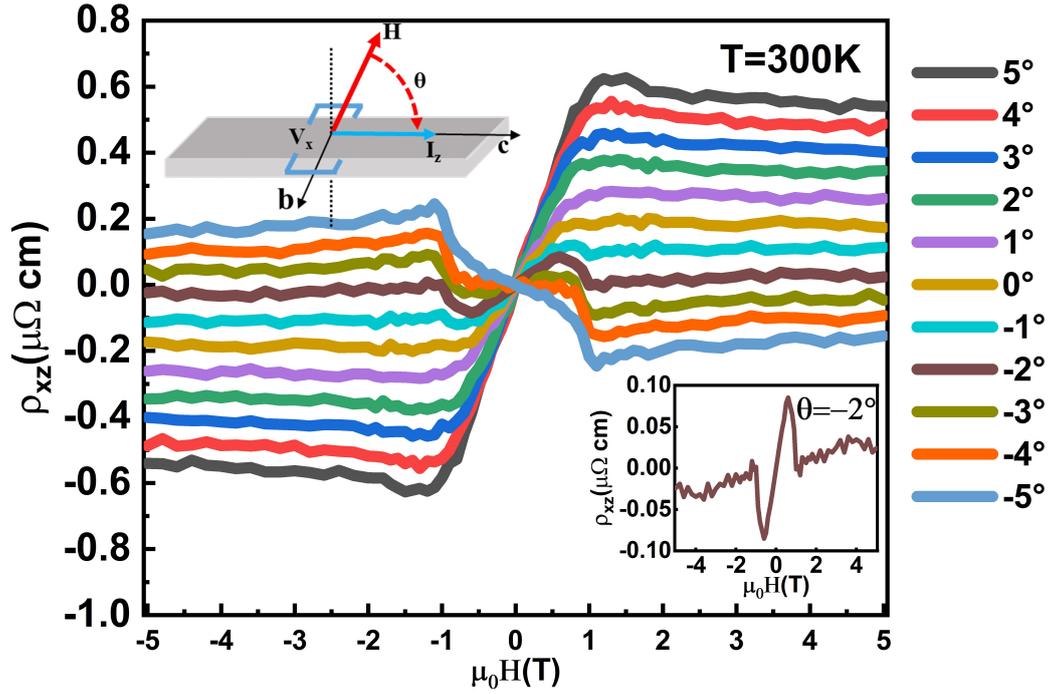

**Fig. S2.** Transverses resistivity $\rho_{xz}$ with $H$ rotated around $b$ axis and $I$ along $c$ axis from $\theta=5°$ to -5° at 300 K. The upper inset is the schematic diagram of measurement and the bottom is the transverses resistivity curve at $\theta=-2°$. All of the transverse resistivity curves show obvious turning point near the saturation field.

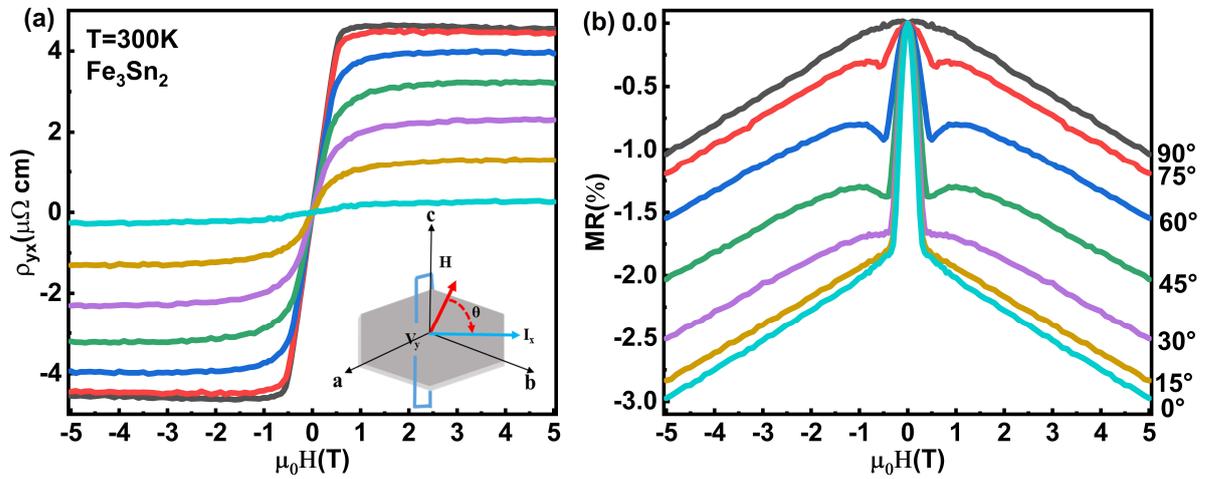

**Fig. S3.** (a) Transverses resistivity $\rho_{yx}$ and (b) magnetoresistance of Fe$_3$Sn$_2$ with $H$ rotated from $c$ axis to $ab$ plane and $I$ along ab plane at 300 K. The inset of (a) is the schematic diagram of measurement.

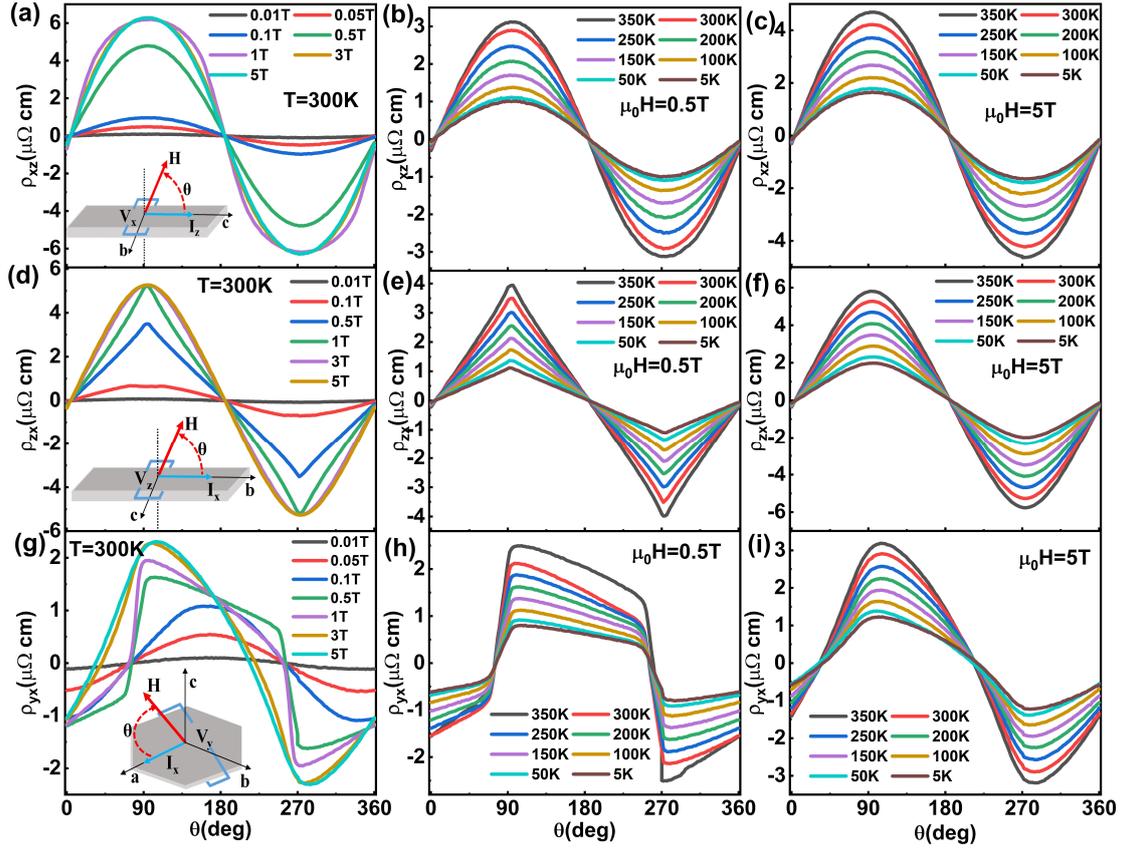

**Fig. S4**. Angles dependent of transverses resistivity of $Fe_5Sn_3$. (a-c) Angles dependent of transverses resistivity with *H* rotated around *b* axis and *I* along *c* axis. (d-f) Angles dependent of transverses resistivity with *H* rotated around *c* axis and *I* along *b* axis. (g-i) Angles dependent of transverses resistivity with *H* rotated from *ab* plane to *c* axis and *I* along *ab* plane.

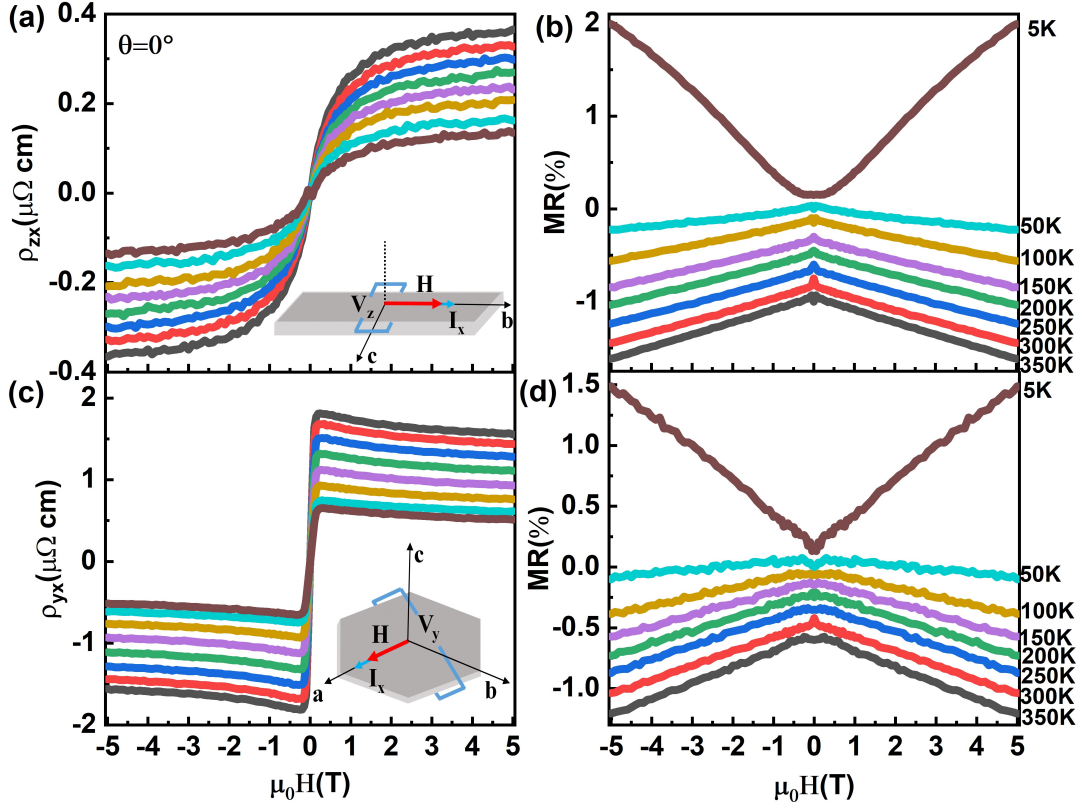

**Fig. S5**. Transverse resistivity and magnetoresistance curves when *H* was parallel to *I* at various temperature.

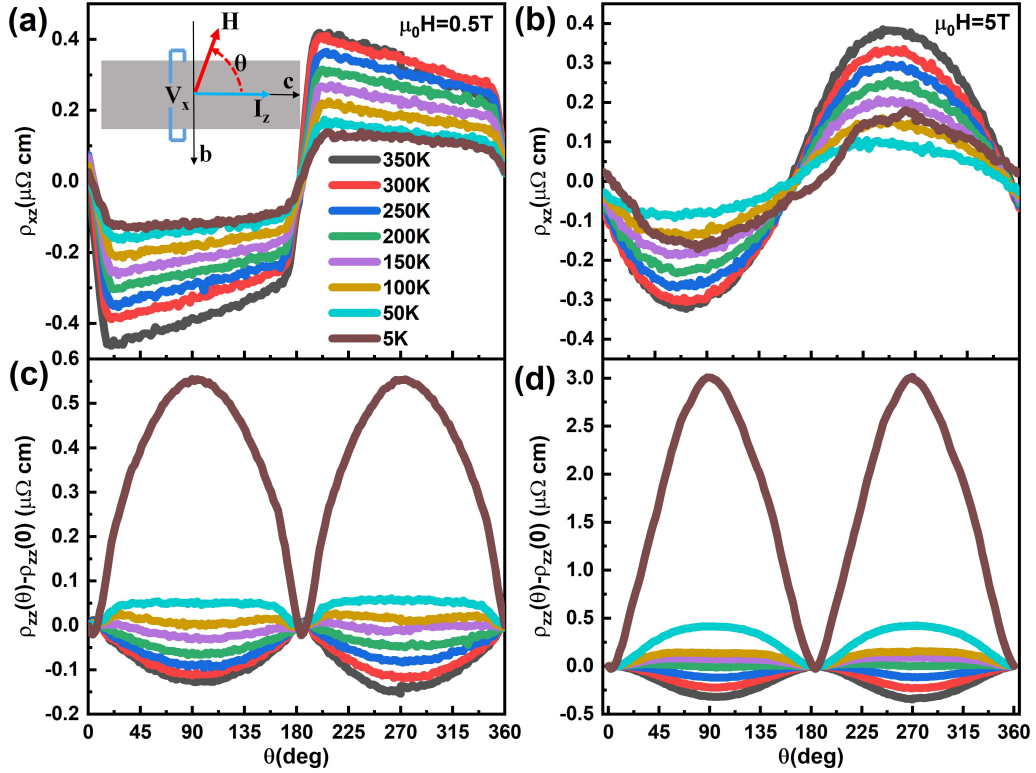

**Fig. S6**. Angles dependent of transverses resistivity and longitudinal resistivity with *H* rotated in *bc* plane and *I* current *c* axis.